\documentstyle[epsfig,aps,twocolumn]{revtex}
\title{Phase-selective reversible quantum decoherence in
cavity QED experiment}
\author{R. Filip\footnote{email:filip@optnw.upol.cz, tel:+420-68-5631572,
fax:+420-68-5224246}\\
Department of Optics, Palack\' y University,\\
17. listopadu 50,  772~07 Olomouc, \\ Czech Republic}
\date{\today}
\begin{document}
\maketitle
\begin{abstract}
New feasible cavity QED experiment is proposed to analyse reversible quantum
decoherence in consequence of quantum complementarity and entanglement.
Utilizing the phase-sensitive manipulations with environment, it is
demonstrated how the complementarity particularly induces a preservation of
visibility, whereas quantum decoherence is more
progressive due to pronounced entanglement
between system and environment.
This effect can be directly observed using the proposed cavity QED
measurements.
\end{abstract}
PACS number(s): 42.50.Dv\newline

\section{Introduction}

Quantum interference is essential phenomenon to
understand the quantum-mechanical effects.
The reason why the quantum interference is hard to observe experimentally,
is decoherence arising from an interaction of the
system with environment. In the last decade, the QED cavity experiments
\cite{QED1,Raimond97}
or experiments with trapped ions \cite{ion} were
performed to investigate the decoherence phenomenon
from point of view reversible and irreversible nature.
The essential principle of decoherence can
be generalized in the following way.
Due to system-environmental interaction, mutual
quantum entanglement generates a stronger correlations
between system end environment than it
can be achieved in classical domain.
Then, if one treats the system separately to
their environment, these strong correlations are reduced
and quantum interference vanishes in the system.

Exhibitions of quantum decoherence are often experimentally observed
in a decreasing of the visibility of interference fringes in
particular measurement.
Thus, the vanishing of quantum interference
is treated as basis-dependent problem.
On the other hand, due to
progressive system-environmental entanglement,
the particular entropy (purity) of the system
substantially increases and fidelity between states before and
after decoherence process decreases.
Contrary, this point of view is independent of a choice of basis and is not
related to any particular measurement.

In this Paper,
based on experimental
arrangement with reversible decoherence \cite{Raimond97},
we propose new feasible cavity QED experiment to analyse the
decoherence from both
points of view: visibility and entanglement, using the directly measurable
quantities.
We can conclude that whereas the visibility
in particular measurement can be preserved by environmental
state squeezing, the state-independent characteristics of decoherence
(entropy, purity, fidelity)
show that the decoherence is more progressive as the visibility is
pronouncedly preserved.
The preserving of visibility
is closely related to the complementarity between visibility and
distinguishability. On the other hand, the pronounced
decoherence observed in the entropy and purity
is connected with a large entanglement between system and
environment. Due to unitary evolution of total system,
the decoherence process can be considered ideally reversible.
However, it is hard to obtain this case in realistic experiment
and therefore the cavity losses and coupling to other modes must be
considered.
To observe both the effects
experimentally, direct measurements of the visibility, negativity of
Wigner function, purity and fidelity are proposed for cavity QED
experiments. 

\section{State preparation and measurement}

The basic experimental scheme is depicted on Fig.~1.
It modifies a previously used experimental arrangement
\cite{Raimond97} to test the reversibility of quantum decoherence.
The arrangement utilizes the Ramsey atomic
interferometer where a two-level atom cross the
high-Q cavity $C_{0}$. This single
circular Rydberg atom is treated to prepare the coherent state
superposition in cavity $C_{0}$. Then the cavity
$C_{0}$ is coupled during an appropriate time to auxillary
resonant cavity $C_{1}$ in the quadrature squeezed state and quantities
of interest are measured using atomic probe.

The principle of the coherent state preparation is
discussed in detail in \cite{QED2}.
It is generated by the dispersive, nonresonant coupling of a
single circular Rydberg atom to
the cavity mode, in which a coherent field $|\xi(0)\rangle$ is
initially prepared. Before the atoms enter the cavity, they are excited
by $\pi/2$-pulse in first Ramsey zone to balanced
superposition of levels $|b\rangle$ and $|c\rangle$. An atom crossing
the cavity in the state $|b\rangle$ induces appreciable phase shift on
the field in the cavity. Let us assume that this shift can be adjusted
to a value exactly equal to $\pi$ by proper selection of the atomic
velocity (about 100m/s). Then a coherent field $|\xi(0)\rangle$ in
cavity is transformed into $|-\xi(0)\rangle$. The phase shift is negligible,
if the atom crosses the cavity in state $|c\rangle$.
The atom crosses centimeter-sized cavity in a time of the order of
$10^{-4}s$, much shorter than the field relaxation time (typically
$10^{-3}-10^{-2}s$), and to the atomic radiative damping time ($3\times
10^{-2} s)$. After the atom has crossed the cavity,
it enters into second Ramsey microwave $\pi/2$ pulse and
after detection of the atom in the state $|c\rangle$
(or $|b\rangle$), the following Schr\" odinger-cat states are conditionally
prepared
\begin{eqnarray}\label{cats}
|\Psi_{\pm}\rangle&=&\frac{1}{\sqrt{N_{\pm}}}
(|\xi(0)\rangle\pm|-\xi(0)\rangle),\nonumber\\
N_{\pm}&=&2\left(1\pm\exp(-2\xi^{2}(0))\right).
\end{eqnarray}
A sign $\pm$ in the superposition can be obtained by postselection in dependence
on output atomic states $|c\rangle$ and $|b\rangle$.

Analogically to experiment \cite{Raimond97},
we consider a controlled resonant coupling preformed by superconducting
waveguide between the
cavities $C_{0}$ and $C_{1}$. In addition, we simplify the situation
by the assuming that the coupling interaction
can be effectively reduced to two modes $S$ and $E$,
representing the system and environment particularly.
Then, the coupling can be
simply described by the following interaction Hamiltonian
$\hat{H}=\hbar\kappa
(\hat{a}_{S}^{\dag}\hat{a}_{E}+\hat{a}_{E}^{\dag}\hat{a}_{S})$,
where $\kappa$ is effective coupling constant
and $\hat{a}_{S}$ and $\hat{a}_{E}$
are annihilation operators of particular modes.
Similarly to experimental arrangement \cite{Raimond97},
this interaction could be tuned by adjusting the cavities coupling through a
superconducting waveguide. We will consider a strongly slowed down the
relaxation processes for the
both cavities (as well as for the atoms during the state preparation).
This assumption is realistic for the quality factor of cavities at least
$Q\approx 10^{9}$ \cite{Raimond97}
(corresponding photon lifetimes are order of a few $ms$).
Let us assume that the coupling between the cavities plays no role
during the cat generation in cavity $C_{0}$. In addition,
the state preparation in cavity $C_{1}$,
provided the preparation times are much shorter
than $1/\kappa$.  In summary, the coupling constant $\kappa$
must chosen such that the inequalities $t_{int}\ll 1/\kappa\ll T$,
where $t_{int}$ is atomic transit
time through the cavity ($\approx 20\mu s$)
and $T$ is photon lifetime ($\approx 1ms$) \cite{Raimond97}.

Whereas in the experiment \cite{Raimond97} the cavity $C_{1}$
was considered in the vacuum state, we will assume that the mode
$E$ in cavity $C_{1}$ is prepared before
coupling in the quadrature squeezing vacuum state $|0,r\rangle$.
This squeezed state can be prepared
by a nonlinear medium coherently performing a reduction of fluctuations in
certain quadrature of field. Nonlinear degenerate parametric process
described by the effective Hamiltonian
$\hat{H}=2\hbar\Gamma (\hat{a}_{1}^{\dag 2}-\hat{a}_{1}^{2})$,
where $\Gamma$ is nonlinear coupling constant and starting from vacuum
state can be appropriated.
The squeezed state is obtained after effective
interaction time $\tau=\frac{r}{2\Gamma}$.
For $r>0$ the fluctuations are reduced in
the $\hat{Q}$ quadrature, as the Schr\" odinger-cat state is
orientated, whereas for $r<0$ in the complementary
quadrature $\langle(\Delta\hat{Q}_{E}(0))^{2}\rangle=1/4\times\exp(-2r)$,
$\langle(\Delta\hat{P}_{E}(0))^{2}\rangle=1/4\times\exp(2r)$.
Squeezing in degenerate parametric amplifier was demonstrated
\cite{Kimble86} using MgO:LiNbO$_{3}$ crystal putting inside a high-Q
cavity and pumped by the second harmonic generation of a 1.06$\mu$m
Nd:YAG laser. The noise level measured by balanced homodyne detector was
larger than 50 percent below the vacuum limit, which is corresponding to
the squeezing parameter $r=2 \Gamma t\approx 1$.
To obtain the considered interaction, we
need to adjust the frequency of light in cavity $C_{0}$, effectively
interacting with atomic probe, with a mean frequency of maximal squeezing
in parametric process.

In QED cavity experiments we are not able to directly measure the
fields in high-Q cavities.
It is only possible to study statistical
properties of the cavity field from the statistical properties of the
auxillary atoms that are crossing the cavity.
To observe the values of the Wigner function in phase-space points,
the direct measurement of the
Wigner function in the optical cavity
was proposed \cite{measur}. We will use a simplified
version of the measurement to utilize the identical
experimental setup as for state preparation.
A probe atom in the excited
state $|c\rangle$ is sent through the system.
After the atom crosses the apparatus, the internal state of the atom is
detected by two field ionization detectors. This experiment is repeated
many times, starting each run with the same field, and probabilities
$p_{b}$ and $p_{c}$ of detecting the probe atom in the state $|b\rangle$
and $|c\rangle$ are determined. From the difference of probabilities,
a value of Wigner function in origin of the phase space can be
determined from this simple relation $W(0)=2(p_{c}-p_{b})$.
An important feature is insensitivity to the
detection efficiency of the atomic counters. However the measurement
accuracy depends on the detector's selectivity, i.e. the ability to
distinguish between two atomic states and the velocity spread of atomic
beam. It will be proposed in next section, how the visibility of
interference, distinguishability and negativity of Wigner function parameters
can be directly measured using this setup.

To observe the purity of state and fidelity between
initial prepared coherent state
superposition and state during evolution the scheme proposed in
\cite{overlap} can be used. Using it we are able to measure the overlap
$\mbox{Tr}\hat{\rho}_{1}\hat{\rho}_{2}$ between separable states in two
cavities by means of their mutual coupling and atomic probe measurement
performed by previously discussed
Ramsey interferometer. The overlap is given by the
visibility of interference of repeatedly performed interferometric
experiments.
To measure the purity and fidelity, we need to prepare two cavities
$C_{0}$ and $C'_{0}$ in the same initial coherent state superposition.
The purity can be measured,
if we perform the proposed experiment immediately in two identical setup and
measure the overlap between cavities $C_{0}$ and $C'_{0}$
evolving in the same state.
On the other hand, to measure the fidelity one cavity is affected
by coupling with cavity $C_{1}$ in proposed experimental setup,
whereas the initial state in second cavity is prevented from any changes.
The suggested measurements offer a possibility
to experimentally observe the decoherence from the both different points of
view, as will be discussed in the following sections.

\section{Evolution of quantum interference}

Quantum interference effects can be well illustrated in behavior of
Wigner function and density matrix in coordinate representation.
After coupling time $t$,
Wigner function corresponding to the particular states of
signal mode $S$ in cavity $C_{0}$
and environmental mode $E$ in cavity $C_{1}$
can be found in the following form
\begin{eqnarray}\label{Wign}
W_{\pm i}(Q,P,t)&=&2N^{-1}_{\pm}\Gamma_{i}(P)\Gamma_{i}(Q)\times\nonumber\\
& &\times\left[O_{i}\cosh\left(\frac{\xi_{i}(t) Q}
{\langle (\Delta\hat{Q}_{i})^{2}\rangle}\right)\pm\right.\nonumber\\
& &\left.\pm R_{i}\cos\left(\frac{\mu_{i}(t) P}
{\langle (\Delta\hat{P}_{i})^{2}\rangle}\right)\right],
\end{eqnarray}
where $i=S,E$ and
$\pm$ corresponds to particular states (\ref{cats}) and
\begin{eqnarray}\label{R}
\Gamma_{i}(P)&=&\frac{1}{\sqrt{2\pi\langle
(\Delta\hat{P}_{i}(t))^{2}\rangle}}\exp\left(-\frac{P^{2}}{2\langle
(\Delta\hat{P}_{i}(t))^{2}\rangle}\right),\nonumber\\
R_{i}&=&\exp\left(-2\xi^{2}\right)\exp\left(\frac{\mu_{i}^{2}(t)}{2\langle
(\Delta\hat{P}_{i}(t))^{2}\rangle}\right),
\end{eqnarray}
\begin{eqnarray}\label{D}
\Gamma_{i}(Q)&=&\frac{1}{\sqrt{2\pi\langle
(\Delta\hat{Q}_{i}(t))^{2}\rangle}}\exp\left(-\frac{Q^{2}}{2\langle
(\Delta\hat{Q}_{i}(t))^{2}\rangle}\right),\nonumber\\
O_{i}&=&\frac{\exp(-2\xi^{2}(0))}{D_{i}}\nonumber\\
D_{i}&=&\exp(-2\xi^{2}(0))\exp\left(-\frac{\xi_{i}^{2}(t)}{2\langle
(\Delta\hat{Q}_{i}(t))^{2}\rangle}\right).
\end{eqnarray}
Density operator of mode in cavity in the coordinate
representation may be calculated as the Fourier transform of the Wigner
function $W(Q,P,t)$ and expressed in the following form
\begin{eqnarray}\label{diag1}
p_{\pm i}(Q,Q')&=&
\frac{1}{N_{\pm}\sqrt{2\pi\langle(\Delta\hat{Q}_{i})^{2}\rangle}}\times\nonumber\\
& &\times\left[\exp\left(-2(\frac{(Q-Q')}{2})^{2}\langle(\Delta\hat{P}_{i})^{2}\rangle
\right)\times\right.\nonumber\\
& &\left.\times\left(\exp\left(-\frac{(\frac{Q+Q'}{2}-
\xi_{i}(t))^{2}}{2\langle(\Delta\hat{Q}_{i})^{2}\rangle}\right)\right.\right.+\nonumber\\
& &\left.\left.+\exp\left(-\frac{(\frac{Q+Q'}{2}+\xi_{i}(t))^{2}}{
2\langle(\Delta\hat{Q}_{i})^{2}\rangle}\right)\right)+\right.\nonumber\\
& &\left.\pm R_{i}\exp\left(-
\frac{(\frac{Q+Q'}{2})^{2}}{2\langle(\Delta\hat{Q}_{i})^{2}\rangle}\right)\right.\times\nonumber\\
& &\left.\times\left(\exp\left(-2(\frac{\mu_{i}(t)}{2\langle(\Delta\hat{P}_{i})^{2}\rangle}+\right.\right.\right.\nonumber\\
& &\left.\left.\left.+\frac{Q-Q'}{2})^{2}\langle(\Delta\hat{P}_{i})^{2}\rangle\right)+\right.\right.\nonumber\\
& &\left.\left.+\exp\left(-2(\frac{\mu_{i}(t)}{2\langle(\Delta\hat{P}_{i})^{2}\rangle}-\right.\right.\right.\nonumber\\
& &\left.\left.\left.-\frac{Q-Q'}{2})^{2}\langle(\Delta\hat{P}_{i})^{2}\rangle\right)\right)\right].
\end{eqnarray}

For the proposed experiment operating with initial squeezed
state in cavity $C_{1}$,
after some calculations incorporating a damping in both the cavities
$C_{0}$ and $C_{1}$, the evolution of particular time-dependent
parameters for both the modes $S,E$ can be expressed in following form
\begin{eqnarray}
\xi_{S}(t)&=&\mu_{S}(t)=\frac{\exp(-\gamma_{+}t/2)}{\lambda}f(t)\xi(0),\nonumber\\
\xi_{E}(t)&=&-\mu_{E}(t)=\frac{\exp(-\gamma_{+}t/2)}{\lambda}g(t)\xi(0),
\end{eqnarray}
\begin{eqnarray}
\langle (\Delta\hat{Q}_{S}(t))^{2}\rangle&=&
\frac{\exp(-\gamma_{+}t)}{\lambda^{2}}\left(f_{+}^{2}(t)\langle
(\Delta\hat{Q}_{S}(0))\rangle+\right.\nonumber\\
& &\left.+g^{2}(t)\langle (\Delta\hat{P}_{E}(0))\rangle\right)+F_{S}(t),\nonumber\\
\langle (\Delta\hat{P}_{S}(t))^{2}\rangle&=&
\frac{\exp(-\gamma_{+}t)}{\lambda^{2}}\left(f_{+}^{2}(t)\langle
(\Delta\hat{P}_{S}(0))\rangle+\right.\nonumber\\
& &\left.+g^{2}(t)\langle
(\Delta\hat{Q}_{E}(0))\rangle\right)+F_{S}(t),\nonumber\\
\langle (\Delta\hat{Q}_{E}(t))^{2}\rangle&=&
\frac{\exp(-\gamma_{+}t)}{\lambda^{2}}\left(f_{-}^{2}(t)\langle
(\Delta\hat{Q}_{E}(0))\rangle+\right.\nonumber\\
& &\left.+g^{2}(t)\langle (\Delta\hat{P}_{S}(0))\rangle\right)+F_{E}(t),\nonumber\\
\langle (\Delta\hat{P}_{E}(t))^{2}\rangle&=&
\frac{\exp(-\gamma_{+}t)}{\lambda^{2}}\left(f_{-}^{2}(t)\langle
(\Delta\hat{P}_{E}(0))\rangle+\right.\nonumber\\
& &\left.+g^{2}(t)\langle (\Delta\hat{Q}_{S}(0))\rangle\right)+F_{E}(t),
\end{eqnarray}
\begin{eqnarray}
f_{\pm}(t)&=&\gamma_{-}\sin\lambda t/2+\lambda\cos\lambda
t/2,\nonumber\\
g(t)&=&2\kappa\sin\lambda t/2,\nonumber\\
F_{S,E}(t)&=&\frac{1}{2}\left[\left(\eta_{S,E}\lambda^{2}(\lambda^{2}+\gamma_{+}^{2}+
\gamma_{-}^{2})+\right.\right.\nonumber\\
& &\left.\left.+4\eta_{E,S}\lambda^{2}\kappa^{2}\right)
\left(1-\exp(-\gamma_{+}t)\right)+\right.\nonumber\\
& &+\left.\eta_{S,E}\lambda^{2}\gamma_{+}(\gamma_{+}+2\gamma_{-})\times\right.\nonumber\\
& &\left.\times\left(
1-\exp(-\gamma_{+}t)\cos\lambda t\right)-\right.\nonumber\\
& &\left.-\gamma_{+}^{2}(\eta_{S,E}\gamma_{-}^{2}+4\eta_{E,S}\kappa^{2})\times\right.\nonumber\\
& &\left.\times\left(1-\cos\lambda t\right)\exp(-\gamma_{+} t)+\right.\nonumber\\
& & \left.\left(\eta_{S,E}\lambda\gamma_{-}(\lambda^{2}-\gamma_{-}^{2}-2\gamma_{+}
\gamma_{-})-\right.\right.\nonumber\\
& &\left.\left.-4\eta_{E,S}\kappa^{2}\lambda\gamma_{+}\right)
(\exp(-\gamma_{+}t)\sin\lambda t\right],\nonumber\\
\lambda &=&\sqrt{4\kappa^{2}-\gamma_{-}^{2}},\hspace{0.3cm}
\gamma_{+}=\gamma_{S}+\gamma_{E},\nonumber\\
\gamma_{-}&=&\gamma_{E,S}-\gamma_{S,E},\hspace{0.3cm}\eta_{S,E}=\gamma_{S,E}(\langle
n_{S,E}\rangle+\frac{1}{2}),\nonumber\\
\langle(\Delta\hat{Q}_{S}(0))^{2}\rangle&=&
\langle(\Delta\hat{P}_{S}(0))^{2}\rangle=1/4,\nonumber\\
\langle(\Delta\hat{Q}_{E}(0))^{2}\rangle&=&1/4\times\exp(-2r),\nonumber\\
\langle(\Delta\hat{P}_{E}(0))^{2}\rangle&=&1/4\times\exp(2r),\nonumber\\
\end{eqnarray}
where $\gamma_{S}$ and $\gamma_{E}$ are damping constant of particular
modes and $\langle n_{S}\rangle$ and $\langle n_{E}\rangle$ are
corresponding mean numbers of thermal reservoir fluctuations.

The decoherence effect exhibits in the decreasing of visibility
in particular field quadrature measurements.
For given coherent state superposition,
particular interference effect can be observed
in marginal distribution of quadrature $\hat{P}$
\begin{equation}\label{probP}
p_{i}(P)=\frac{2}{N_{\pm}}\Gamma_{i}(P)\left[1\pm R_{i}
\cos\left(\frac{\mu_{i}P}{\langle (\Delta\hat{P}_{i})^{2}\rangle}\right)
\right],
\end{equation}
where $i=S,E$. This equation represents an analogue of interference rule in
any point of phase space. To measure the visibility $R_{i}$,
we can consider probability $p_{i}(0)$ in origin of distribution for
both the states $|\Psi_{\pm}\langle$.
Then interference formula (\ref{probP}) can be simply expressed
in following form $p_{\pm}(0)=2/N_{\pm}(1\pm R_{i})$,
where $R_{i}$ is visibility defined in the following way
\begin{equation}\label{visib}
R=\frac{N_{+}p_{+}(0)-N_{-}p_{-}(0)}{N_{+}p_{+}(0)+N_{-}p_{-}(0)}.
\end{equation}
The normalization factors $N_{\pm}$ must be introduced in definition of
visibility due to
a non-orthogonality of coherent states. They can be approximately considered
as unity if amplitude $\xi(0)$ is sufficiently large ($\xi(0)>2$).
Then Eq.~(\ref{visib}) approaches commonly used definition of visibility.

Immediately, the marginal distribution for $\hat{Q}$ quadrature
\begin{eqnarray}\label{probQ}
p_{i}(Q)&=&\frac{2}{N}\exp(-2\xi^{2}(0))\Gamma_{i}(Q)\times\nonumber\\
& &\times\left[D^{-1}_{i}\cosh\left(\frac{\xi_{i}Q}{\langle
(\Delta\hat{Q}_{i})^{2}\rangle}\right)\pm 1\right]
\end{eqnarray}
consists (for sufficiently large $\xi(0)$) essentially of two
Gaussian peaks symmetrically distant from the origin.
It can be evident from another form of Eq.~(\ref{probQ})
\begin{equation}
p_{i}(Q)=
\frac{1}{N}\left(p_{L}(Q)+p_{R}(Q)\pm 2\exp\left(-2\xi^{2}(0)
\right)\Gamma_{i}(Q)\right),
\end{equation}
where
\begin{equation}
p_{L,R}(Q)=\frac{1}{\sqrt{2\pi\langle
(\Delta\hat{Q}_{i})^{2}\rangle}}\exp\left(-\frac{(Q\pm\xi_{i})^{2}}{
2\langle (\Delta\hat{Q}_{i})^{2}\rangle}\right)
\end{equation}
is probability distribution of the marginal peaks and $\Gamma_{i}(Q)$ represents the
vacuum noise influence. Analogically to interference parameter $R_{i}$,
the overlap parameter $O_{i}$
can be proposed. It can be defined as square root of normalized overlap
between separated marginal probability distributions corresponding to
particular initial coherent states $|\alpha\rangle$ and
$|-\alpha\rangle$
\begin{equation}
O^2=\frac{\int_{-\infty}^{\infty}p_{L}(Q)p_{R}(Q)dQ}{
\int_{-\infty}^{\infty}p_{L0}(Q)p_{R0}(Q)dQ},
\end{equation}
where $\exp(-2\xi^{2}(0))\leq O~\leq 1$ and
$p_{L0}(Q)=p_{R0}(Q)\equiv\Gamma(Q)$ are the considered Gaussian
distributions $p_{L}(Q)$, $p_{R}(Q)$ shifted to origin and having
the same variance. The visibility $R_{i}$ and overlaps
$O_{i}$ are connected by a relation $R_{S}R_{E}\leq O_{S}O_{E}$, where
equality occurs for unitary evolution of total system.
In cavity QED experiments, the visibility $R_{i}$ and overlap
$O_{i}$ parameters can be straightforwardly calculated from
the measurement of Wigner function $W_{i}(0,0)$
in origin of phase space
\begin{eqnarray}\label{RO}
R&=&\frac{N_{+}W_{+}(0,0)-N_{-}W_{-}(0,0)}{4W_{0}(0,0)},\nonumber\\
O&=&\frac{N_{+}W_{+}(0,0)+N_{-}W_{-}(0,0)}{4W_{0}(0,0)},
\end{eqnarray}
where $W_{0}$ is Wigner function of the particular mode for initial
vacuum state. A value of Wigner
function can be directly determined by atomic-probe measurement as was referred in
the previous section. If $\xi(0)$ is sufficiently large ($\xi(0)>2$), then
the relations (\ref{RO}) can be simplified to
$R\approx (W_{+}(0,0)-W_{-}(0,0))/2W_{0}(0,0)$ and
$O\approx (W_{+}(0,0)+W_{-}(0,0))/2W_{0}(0,0)$.

To simply describe a distinguishability between marginal
peaks, the parameters $D_{i}$ (\ref{D}) inversely related
to overlap $O_{i}$ can be used.
This choice of distinguishability parameter is appropriate
for homodyne measurement of particular quadrature operators, where
both the parameters $D_{i}$ and $R_{i}$ can be determined from values of
marginal probability distribution of $\hat{Q}$ and $\hat{P}$ operators
in the origin, using (\ref{visib}) in both the particular cases.
In cavity QED experiments, the distinguishability parameter $D_{I}$ can be
determined from overlap $O_{i}$.
The distinguishability and
interference parameters in the particular modes
are connected by the relation
\begin{equation}\label{comple}
R_{S}D_{S}R_{E}D_{E}\leq\exp(-4\xi^{2}(0)),
\end{equation}
where equality is achieved for unitary evolution of system and
environmental modes.
Particularly, the products $R_{i}D_{i}$ satisfy the following inequalities
\begin{equation}
\exp\left(-4\xi^{2}\right)\leq R_{E}D_{E}
\leq\exp\left(-2\xi^{2}\right)\leq R_{S}D_{S}\leq 1.
\end{equation}
Another definition of distinguishability parameter leading to different
form of relation between visibility and distinguishability is proposed
in Appendix.

It can be simply found that the first part of the right side of Wigner
function in Eq.~(\ref{Wign})
is connected with distinguishable peaks,
and is always positive, whereas the second part
is connected to interference oscillations in phase space, and can be negative.
In a particular point of phase space, the negative part can be larger
and Wigner function becomes negative.
An occurring of these negative values depends only on sign of expression in
square brackets in Eq.~(\ref{Wign}).
For initial state $|\Psi_{+}\rangle$, the expression in square brackets has
the maximal negative value in the points
$(0,\frac{\pi\langle\Delta\hat{P}_{i})^{2}\rangle}{\mu_{i}})$
of the phase space.
After some calculations,
the necessary and sufficient condition for the occurring of a
negative value of the Wigner function can be found in the
following form
\begin{equation}
D_{i}R_{i}>\exp(-2\xi^{2}(0))
\end{equation}
for the particular mode $i=S,E$. A positive Wigner function can be
treated semiclassically,
whereas for Wigner function with the negative values no semiclassical
theory can be constructed.
Irrespective to,
the Glauber-Sudarshan quasidistribution cannot be positive or exist
everyone. To measure product $D_{i}R_{i}$, the
atomic-probe measurement of Wigner function can be used. We define
a new parameter $N_{i}=D_{i}R_{i}\exp(2\xi^{2}(0))$ and it can be
determined from the values of Wigner functions in origin of phase space
by the following relation
\begin{equation}
N=\frac{N_{+}W_{+}(0,0)-N_{-}W_{-}(0,0)}
{N_{+}W_{+}(0,0)+N_{-}W_{-}(0,0)}.
\end{equation}
As negative values of Wigner function
are suppressed, the parameter $N_{i}$ decreases to unity.
For sufficiently large $\xi(0)$
($\xi(0)>2$), the normalization factors can be approximately considered as
unity and thus the relation can be simplified.
The negativity parameters $N_{i}$ satisfy simple relation
$N_{S}N_{E}\leq 1$, which describe exchange of nonclassicality
between the system and environment mode.
Assuming unitary
evolution of the total system, the relation
(\ref{comple}) can be rewritten for system mode to the following form
\begin{equation}\label{comple2}
R_{S}D_{S}=\exp(-2\xi^{2}(0))N_{S}.
\end{equation}
Thus, a complementarity between visibility and distinguishability
is sharper as the nonclassical character of the state vanishes.

Quantum decoherence process is assisted by an entanglement between system and
environment. As the entanglement increases, the state of the system
becomes more mixed. Therefore as an appropriate measurable degree of
quantum decoherence, the purity of particular system state can be used.
In addition, if the total system is in a pure
state, then the purity can be immediately used as the measure of
entanglement. The purity parameter $P=\mbox{Tr}\hat{\rho}^{2}(t)$
of the state can be expressed by means of
the Wigner function $W(Q,P)$ and for initial state
$|\Psi_{+}\rangle$ is given by
\begin{eqnarray}
P_{i}&=&\pi\int\!\!\!\!\int_{-\infty}^{\infty}
W_{+S}^{2}(Q,P,t)dQdP=\nonumber\\
&=&\frac{1}{N^{2}\sqrt{\langle (\Delta\hat{Q}_{i}(t))^{2}\rangle
\langle (\Delta\hat{P}_{i}(t))^{2}\rangle}}
\Bigl[
1+R^{2}_{i}(t)+\nonumber\\
& &
+\exp(-4\xi^{2}(0))\left(1+\frac{1}{D^{2}_{i}(t)}\right)+\nonumber\\
& &+2\exp(-2\xi^{2}(0))\sqrt{\frac{R_{i}(t)}{D_{i}(t)}}\Bigr].
\end{eqnarray}
From a value of purity, the Renyi entropy $S_{i}=-\ln P_{i}$
\cite{Horodecki96} can be simply calculated.
If state of total system is pure in the initial time and
mutual interaction is unitary, the decreasing
of purity in particular mode is exhibition of pronounced
quantum entanglement and immediately leads to intensive decoherence.
Note, for the total pure state case, the particular
purity parameters are identical for both the signal and idler modes.
To compare the initial state superposition $|\Psi_{+}\rangle$
with decohered state $\hat{\rho}$, the
fidelity $F_{i}=\langle\Psi_{+}|\hat{\rho}_{i}(t)|\Psi_{+}\rangle$ can
be used. It can be calculated form the Wigner function in the following form
\begin{eqnarray}
F_{i}&=&\pi\int\!\!\!\!\int_{-\infty}^{\infty}
W_{i}(Q,P,0)W_{i}(Q,P,t)dQdP=\nonumber\\
&=&\frac{2N^{-2}}{\sqrt{(\langle(\Delta\hat{Q}_{i}(t))^{2}\rangle+\frac{1}{4})
(\langle(\Delta\hat{P}_{i}(t))^{2}\rangle+\frac{1}{4})}}\times\nonumber\\
& &\times\left(\frac{\exp(-4\xi^{2}(0))}
{2D_{i}(t)}\times\right.\nonumber\\
& &\left.\times\left(\exp\left(\frac{2\left(\langle(\Delta\hat{Q}_{i}(t))^{2}\rangle\xi^{2}(0)+
\frac{\xi_{i}(t)}{4}\right)^{2}}{\langle(\Delta\hat{Q}_{i}(t))^{2}\rangle
\left(\langle(\Delta\hat{Q}_{i}(t))^{2}\rangle+\frac{1}{4}\right)}\right)\right.\right.+\nonumber\\
& &\left.\left.+\exp\left(\frac{2\left(\langle(\Delta\hat{Q}_{i}(t))^{2}\rangle\xi^{2}(0)-
\frac{\xi_{i}(t)}{4}\right)^{2}}{\langle(\Delta\hat{Q}_{i}(t))^{2}\rangle
\left(\langle(\Delta\hat{Q}_{i}(t))^{2}\rangle+\frac{1}{4}\right)}\right)\right)+\right.\nonumber\\
&
&\left.\times\frac{R}{2}\left(\exp\left(-\frac{2\left(\langle(\Delta\hat{P}_{i}(t))^{2}\rangle\xi^{2}(0)+
\frac{\mu_{i}(t)}{4}\right)^{2}}{\langle(\Delta\hat{P}_{i}(t))^{2}\rangle
\left(\langle(\Delta\hat{P}_{i}(t))^{2}\rangle+\frac{1}{4}\right)}\right)\right.\right.+\nonumber\\
& &\left.\left.+\exp\left(-\frac{2\left(\langle(\Delta\hat{P}_{i}(t))^{2}\rangle\xi^{2}(0)-
\frac{\mu_{i}(t)}{4}\right)^{2}}{\langle(\Delta\hat{P}_{i}(t))^{2}\rangle
\left(\langle(\Delta\hat{P}_{i}(t))^{2}\rangle+\frac{1}{4}\right)}\right)\right)+\right.\nonumber\\
& &\left.+R_{i}(t)\exp\left(-2\xi^{2}(0)\right)\exp\left(\frac{2\xi^{2}(0)
\langle(\Delta\hat{Q}_{i}(t))^{2}\rangle}
{\left(\langle(\Delta\hat{Q}_{i}(t))^{2}\rangle+\frac{1}{4}\right)}\right)
\times\right.\nonumber\\
& &\left.\times\exp\left(\frac{-\mu^{2}_{i}(t)}
{8\langle(\Delta\hat{P}_{i}(t))^{2}\rangle
\left(\langle(\Delta\hat{P}_{i}(t))^{2}\rangle+\frac{1}{4}\right)}\right)\right.+\nonumber\\
& &\left.+\frac{\exp\left(-2\xi^{2}(0)\right)}{D_{i}}\exp\left(-\frac{2\xi^{2}(0)
\langle(\Delta\hat{P}_{i}(t))^{2}\rangle}
{\left(\langle(\Delta\hat{P}_{i}(t))^{2}\rangle+\frac{1}{4}\right)}\right)\right.\times\nonumber\\
& &\times\left.\exp\left(\frac{\xi_{i}^{2}(t)}
{8\langle(\Delta\hat{Q}_{i}(t))^{2}\rangle
\left(\langle(\Delta\hat{Q}_{i}(t))^{2}\rangle+\frac{1}{4}\right)}\right)\right).
\end{eqnarray}
As fidelity decreases, the state in cavity
mode is far from initial coherent state superposition.
Note that fidelity between initial coherent state superposition and
corresponding mix of coherent states is
$F=1/2(1+\exp(-2\xi^{2}(0)))$. The purity and fidelity parameters can be
directly measured, as has been pointed out in the previous section.

\section{Complementarity and entanglement in decoherence process}

In this section, we will analyse the quantum decoherence from point of
view the complementarity and quantum entanglement in dependence on
squeezing of environmental mode.
In order to keep the discussion as simple as possible, we assume
now the coupling between cavities without an influence of damping
($\gamma_{S}=\gamma_{E}=0$).
System mode $S$ is prepared in coherent state superposition
$|\Psi_{+}\rangle$ and
squeezed state $|0,r\rangle$ is generated in environmental mode $E$
before coupling with the system. After
coupling time rescaled to $G=\kappa t$, decay of
the visibility $R_{S}$ in the system mode can be slowed down, if the
squeezing $r$ is sufficiently positive. The decay
exhibits a non-exponential character with pronounced offset, as can be
seen in Fig.~4(a).
Thus appropriate squeezing vacuum state prepared in environment
can lead to qualitative change of the
exponential to nonexponential vanishing of visibility.
This effect was previously mentioned
\cite{squeez1}
in efficient homodyne detection and in works
\cite{squeez2}
related to phase-sensitive damping and amplification.
As amplitude of coherent state $\xi(0)$ increases, still large squeezing
of vacuum fluctuations is needed for a visibility preserving.

Let us now more precisely
analyse this slowing down of decoherence
using the measurable quantities in proposed experiment.
First, it can be found that distinguishability $D_{S}$ in complementary
quadrature is immediately quickly vanished, as can be observed in Fig.~2.
This effect looks particularly simply when the notion of
quantum complementarity is considered:
due to enhanced noise in $\hat{Q}_{E}$
quadrature ($r$ is sufficiently positive),
the coherent peaks in system mode become more indistinguishable and
immediately, due to relation
between distinguishability and interference (\ref{comple2}),
the visibility is preserved in the measured quadrature $\hat{P}_{S}$.
Thus an indistinguishability of superposed coherent states,
introduced by noise enhancement, can lead to preservation of
visibility in particular measurement.
The complementary situation occurs for $r$ negative,
only with a mutually exchange $R_{S}\leftrightarrow D_{S}$.
For a region near to $r=0$, both
the distinguishability and interference parameters are vanishing,
as was previously experimentally analysed \cite{Raimond97}.

Irrespective to preserving of the visibility, the state is strongly
different from the initial coherent state superposition
due to loss of distinguishable peaks and the
non-classical negative nature of the Wigner function,
as it can be seen in Fig.~3(a,b).
From Fig.~4(b), one can find that the suppression of negative values
is independent of the sign of squeezing parameter $r$.
However, it can be simply found
that Glauber-Sudarshan distribution is still not positive and thus this
interference cannot be treated classically.
In the corresponding density matrix $p(Q,Q')$ in coordinate representation
the off-diagonal peaks are reduced only to half of initial height and is not
completely vanished, as can be seen in Fig.3(d).
This is residuum of quantum interference which is sufficient to generate
the same visibility as for coherent state superposition.
The diagonal peaks
become indistinguishable due to appropriate enhanced noise in
environmental mode. Thus the interference in quadrature $\hat{P}$
is preserved for sufficiently negative $r$,
but some signatures of quantum interference are vanished or degraded.
It can lead to two different levels of interference effect: first is connected
with pronounced negative character of Wigner function and distinguishable
peaks and second with vanishing negative character and undistinguishable
peaks corresponding to particular coherent states.

As can be seen in Fig.~4(c),
an increase of squeezing in the environmental mode leads to
purity vanishing in the system independently
of the directions of squeezing.
Whereas the visibility is preserved for a long time, if $r$ is more
negative, the state in cavity $C_{0}$ is more mixed. Immediately, it is
a signature of pronounced entanglement between cavities,
thank to the total system is in pure state.
Due to pronounced decreasing for large positive $r$,
strong arising entanglement is signature of quantum decoherence,
in contrary to the conjecture obtained from the visibility preserving.
However, the slowest decreasing of purity
is not found for $r=0$, but it is slightly shifted in positive $r$ region.
Thus for considered superposition with $\xi(0)=2$,
the squeezing parameter $r\approx 1$ in environmental mode can
lead to partial slowing down (not pronounced) of entanglement
between system and environmental modes.
The effect of decoherence can be
described by time evolution of fidelity between initial coherent
state superposition and the actual state, which exhibits similar
nonexponential behavior as for the purity (Fig.~4(d)).
Similarly to evolution of purity, the fidelity decreasing can be
slightly slowed down if the squeezing parameter $r\approx 1$, however, the
fidelity is vanished for sufficiently enhanced positive $r$.

In ideal case,
due to unitary evolution of total system and environment, the decoherence
effects are reversible with period $G=\pi$, as was shown in
\cite{Raimond97}. In realistic experiment,
we must consider exponential damping in cavity $C_{0}$ which leads
to degradation of effect of preserving visibility,
as can be seen in Fig.~5(a) for $\gamma_{S}=0.05$ and $\langle
n_{S}\rangle=0$. Naturally, the damping can be induced by the continuum of
modes in cavity $C_{1}$ which are weakly coupled to system mode $S$ in
cavity $C_{0}$.
Whereas for visibility $R$ the degradation is strong for $G<\pi/4$, for
the other parameters $RD$, $P$ and $F$ of decoherence, the difference
to unitary case is only small. In course of time, all the parameters are
vanished, how can be expected.
However, the analysis of
damping case shows that the proposed experiment is feasible due to
contemporary high-Q cavity arrangements.
Note, that for same
damping in cavity $C_{1}$, the exhibitions in particular parameters
are very similar, except for slowly decreasing of the visibility $R$.

\section{Conclusion}

In this paper, a feasible cavity QED experiment was proposed to
analyse quantum decoherence nature using squeezing of
fluctuations in the environment. The results of experiments can be
formulated in terms directly measurable quantities and were analysed
from a different view of quantum decoherence.
Irrespective to
different nature of decoherence process, distinction between particular
preserving of visibility and pronounced quantum decoherence can be observed.
The reason is that the quantum decoherence is rather connected with
degree of entanglement than only with vanishing of visibility in particular
measurement. Therefore the quantum coherence is vanishing irrespective
to behavior in particular quadrature. It is connected with vanishing of
nonclassical negative character of Wigner function in the system as nonclassicality of
the state is transformed to entanglement between system and environment.
On the other hand, the preserving of visibility is
immediately connected with pronounced indistinguishablity in
complementary variable, which
can be analysed using the complementarity principle in the experiments.
Thus, the quantum mechanical complementarity and entanglement
immediately exhibit in phase-sensitive decoherence process causing the
different observable effects.

\medskip
\noindent {\bf Acknowledgments}
The author would like to thank Prof. J. Pe\v rina and J. Fiur\' a\v sek
for discussions stimulating this work.
These results were supported by the project LN00A015 and CEZ:J14/98
of the Ministry of Education of Czech Republic.

\section{Appendix}

The distinguishability parameter can be also defined in a different way
as $D=\sqrt{1-O^{2}}$ from measurable overlap $O$. Using this definition,
it is fixed that $D^{2}+O^{2}=1$ and relations between visibility
$R_{i}$ and
distinguishability $D_{i}$ can be found, if the total system evolves
unitarily, in the following
form similar to commonly used visibility-distinguishability relations
\begin{equation}
\frac{R^{2}_{i}}{N^{2}_{i}}+D_{i}^{2}=1,\hspace{0.3cm}
N_{S}^{2}N_{E}^{2}=1,
\end{equation}
where $i=S,E$ and $N_{i}$ is a parameter of negativity of
Wigner function.
In the analysed cases, it is only matter of convenience
which kind of distinguishability definition is used for the
description.


\newpage
\thispagestyle{empty}
\begin{figure}
\caption{The proposed experimental setup: O -- oven, V -- velocity
selection, B -- excitation box, R$_{1}$, R$_{2}$ -- Ramsey zones,
C$_{0}$ and C$_{1}$ -- high-Q cavities, D$_{e}$, D$_{g}$ ionization
detectores. The atoms are initially excited into upper
Rydberg state by a laser. They cross the cavity between two Ramsey
microwave fields R$_{1}$ and R$_{2}$. After that, an ionization
detector determines the state of the atom.
The cavity C$_{0}$ is
coupled by the superconducting waveguide to cavity C$_{1}$,
where the field quadrature squeezing is performed.}
\end{figure}

\begin{figure}
\caption{Initial squeezed state - complementarity between visibility and distinguishability
parameters $R_{S}$, $D_{S}$, $R_{E}$ and $D_{E}$
in course of rescaled time $G=\kappa t$: (A) $\gamma_{0}=0$, (B)
$\gamma_{0}=0.05$; $\xi(0)=2$, $r=2$, $\gamma_{1}=0$, $n_{1}=n_{0}=0$.}
\end{figure}

\begin{figure}
\caption{Initial squeezed state - decoherence-free case: Wigner function
$W_{+S}(Q,P)$ and density
matrix elements $p_{+S}(Q,Q')$ in coordinate representation
for (a,c) quantum interference ($G=0$) and (b,d) semiclassical
interference exhibitions ($G=\frac{\pi}{4}$); $r=2$, $\xi(0)=2$,
$\gamma_{0}=\gamma_{1}=0$, $n_{1}=n_{0}=0$.}
\end{figure}

\begin{figure}
\caption{Initial squeezed state - decoherence-free case: (a)
visibility $R$, (b) nonclassicality parameter $RD$, (c) purity $P$ and
(d) fidelity
$F$ of internal cavity $C_{0}$ state in dependence on squeezing
parameter $r$ in course of rescaled time $G=\kappa t$; $\xi(0)=2$,
$\gamma_{0}=\gamma_{1}=0$, $n_{1}=n_{0}=0$.}
\end{figure}

\begin{figure}
\caption{Initial squeezed state - decoherence in cavity $C_{0}$:
(a) visibility $R$, (b) nonclasicality parameter $RD$, (c) purity $P$
and (d) fidelity
$F$ of internal cavity $C_{0}$ state in dependence on squeezing
parameter $r$ in course of rescaled time $G=\kappa t$; $\xi(0)=2$,
$\gamma_{0}=0.05$ $\gamma_{1}=0$, $n_{1}=n_{0}=0$.}
\end{figure}


\begin{thebibliography}{99}

\bibitem{QED1}
M. Brune, E. Hagley, J. Dreyer, X. Maitre, A. Maali, C. Wunderlich, J.M.
Raimond, and S. Haroche, Phys.\ Rev.\ Lett.\ {\bf 77}, 4887 (1996);
X. Maitre, E. Hagley, J. Dreyer, A. Maali, C. Wunderlich, M. Brune, J.M.
Raimond, and S. Haroche, J.\ Mod.\ Opt.\ {\bf 44} 2023 (1997).

\bibitem{Raimond97}
J.M. Raimond, M. Brune, and S. Haroche, Phys.\ Rev.\ Lett.\ {\bf 79}
1964 (1997).

\bibitem{ion}
C.J. Myatt, B.E. King, Q.A. Turchette, C.A. Sackett, D. Kielpinski,
W.M. Itano, C. Monroe, and D.J. Wineland, J.\ Mod.\ Opt.\ {\bf 47}
2181 (2000);
Q.A. Turchette, C.J. Myatt, B.E. King, C.A. Sackett, D. Kielpinski,
W.M. Itano, C. Monroe, and D.J. Wineland, Phys.\ Rev.\ A {\bf 62}
053807 (2000).

\bibitem{QED2}
M. Brune, S. Haroche, V. Lefevre, J.M. Raimond, and N.Zagury, Phys.\ 
Rev.\ Lett.\ {\bf 65}, 976 (1990);
M. Brune, S. Haroche, J.M. Raimond, L. Davidovich and N. Zagury,
Phys.\ Rev.\ A {\bf 45} 5193 (1992);
S. Haroche and J.M. Raimond, in {\sl Cavity Electrodynamics}, edited by P.
Berman, Academic Press, New York, (1994);
M. Brune, P. Nussenzveig, F. Schmidt-Kaler, F. Bernardot, A. Maali, J.M.
Raimond and Haroche, Phys.\ Rev.\ Lett.\ {\bf 72}, 3339 (1994);
L. Davidovich, M. Brune, J.M. Raimond, and S. Haroche, Phys.\ Rev.\ A
{\bf 53}, 1295 (1996).

\bibitem{Kimble86}
L-A. Wu, H.J. Kimble, J.L. Hall, H. Wu, Phys.\ Rev.\ Lett.\ {\bf 86}, 
2520 (1986).

\bibitem{measur}
L.G. Lutterbach, and L. Davidovich, Phys.\ Rev.\ Lett.\ {\bf 97}
2546 (1997);
M.S. Kim, G. Antesberger, C.T. Bodendorf and H. Walther, Phys.\ Rev.\
A {\bf 58}, R65 (1998).

\bibitem{overlap}
R. Filip, arXiv:quant-ph/0108119 (2001).

\bibitem{Horodecki96}
R. Horodecki, P. Horodecki, and M. Horodecki, Phys.\ Lett.\ A {\bf 210},
377 (1996).

\bibitem{squeez1}
A. Mecozzi, and P. Tombesi, Phys. \ Lett.\ A {\bf 87}, 101 (1987);
P. Tombesi, and A. Mecozzi, J.\ Opt.\ Soc.\ Am.\ {bf 4}, 1700 (1987);
S.L. Braunstein, Phys.\ Rev.\ A {\bf 45}, 6803 (1992);
U. Leonhardt, Phys.\ Rev.\ A {\bf 48}, 3265 (1993);
U. Leonhardt, Phys.\ Rev.\ A {\bf 49}, 1231 (1993);

\bibitem{squeez2}
M.S. Kim, and V. Bu\v zek, Phys.\ Rev.\ A {\bf 47} 610 (1993);
V. Bu\v zek, M.S. Kim, and Ts. Gantsog, Phys.\ Rev.\ A {\bf 48}, 3394
(1993); R. Filip, J. Opt. B.: Quantum Semiclass. Opt. {\bf 3} 1 (2001);
R. Filip, and J. Pe\v rina, J. Opt. B.: Quantum Semiclass. Opt. {\bf 3}
21 (2001).





\end{thebibliography}
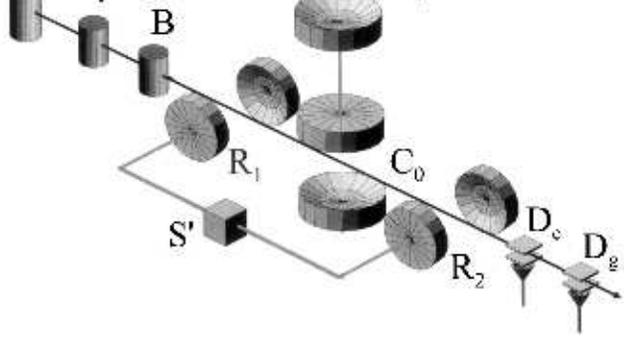
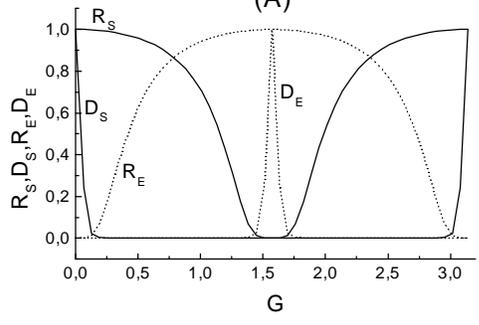
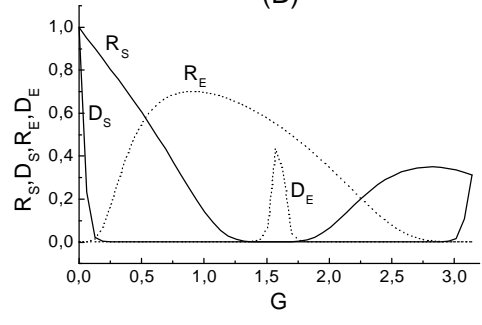
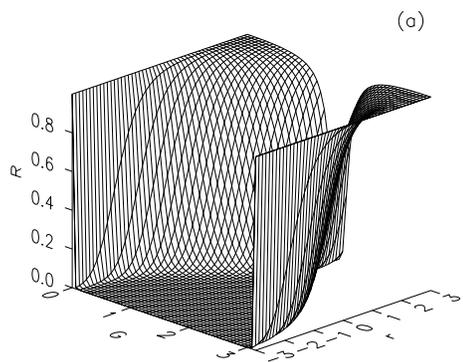
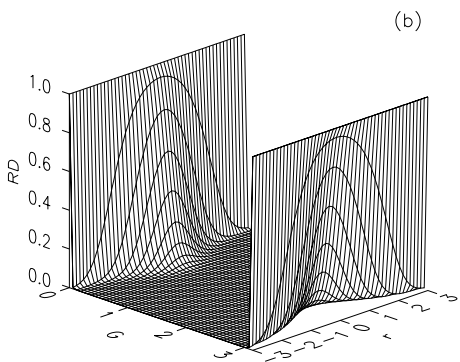
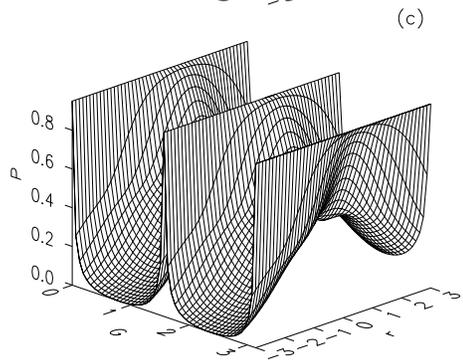
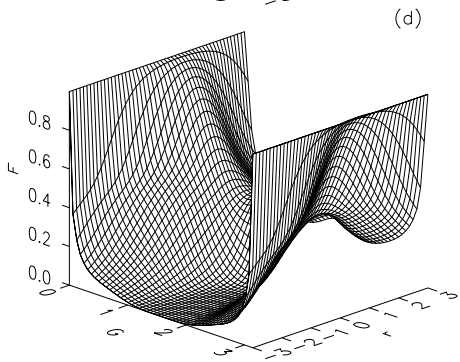
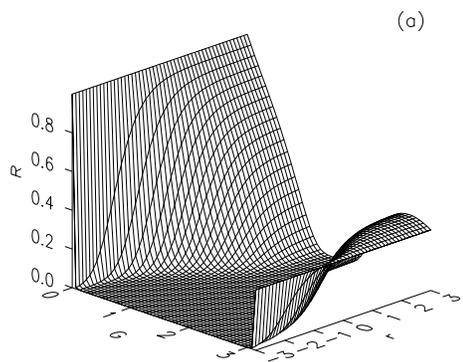
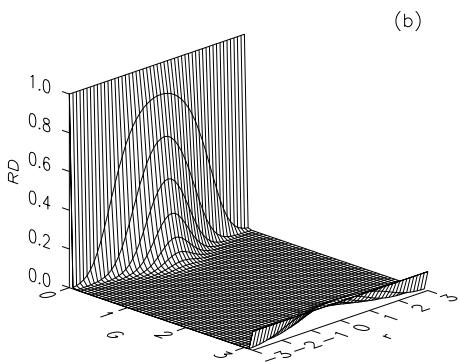
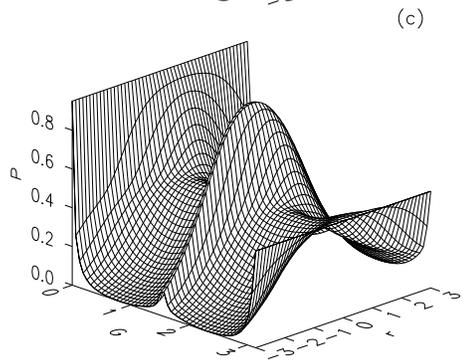
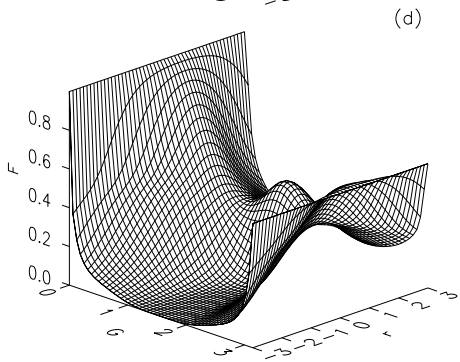
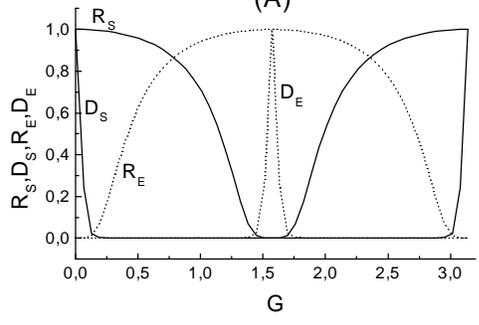
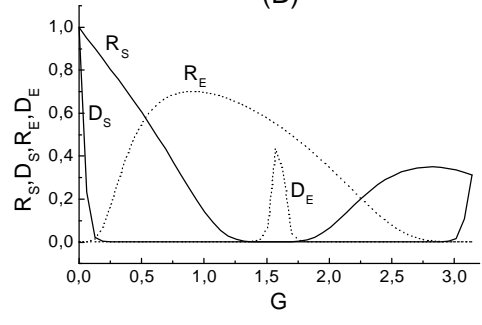
\end{document}